\documentclass[aip,reprint]{revtex4-1}

\usepackage{graphicx}% Include figure files
\usepackage{amsmath}
\usepackage{textcomp} %texterweiterungen
\usepackage{xcolor}
\usepackage{ulem}
\usepackage{amssymb}

\begin{document}

\title{All-electrical detection of spin dynamics in magnetic antidot lattices by the inverse spin Hall effect}

\author{Matthias~B.~Jungfleisch}
\email{jungfleisch@anl.gov}
\affiliation{Materials Science Division, Argonne National Laboratory, Argonne, Illinois 60439, USA}

\author{Wei~Zhang}
\affiliation{Materials Science Division, Argonne National Laboratory, Argonne, Illinois 60439, USA}

\author{{Junjia~Ding}}
\affiliation{Materials Science Division, Argonne National Laboratory, Argonne, Illinois 60439, USA}

\author{Wanjun~Jiang}
\affiliation{Materials Science Division, Argonne National Laboratory, Argonne, Illinois 60439, USA}

\author{Joseph~Sklenar}
\affiliation{Materials Science Division, Argonne National Laboratory, Argonne, Illinois 60439, USA}
\affiliation{Department of Physics and Astronomy, Northwestern University, Evanston, Illinois 60208, USA}

%\author{Frank~Y.~Fradin}
%\affiliation{Materials Science Division, Argonne National Laboratory, Argonne, Illinois 60439, USA}
%\affiliation{Department of Physics and Astronomy, Northwestern University, Evanston, Illinois 60208, USA}

\author{John~E.~Pearson}
\affiliation{Materials Science Division, Argonne National Laboratory, Argonne, Illinois 60439, USA}

\author{John~B.~Ketterson}
\affiliation{Department of Physics and Astronomy, Northwestern University, Evanston, Illinois 60208, USA}

\author{Axel~Hoffmann}
\affiliation{Materials Science Division, Argonne National Laboratory, Argonne, Illinois 60439, USA}

\date{\today}

\begin{abstract}

The understanding of spin dynamics in laterally confined structures on sub-micron length scales has become a significant aspect of the development of novel magnetic storage technologies. Numerous ferromagnetic resonance measurements, optical characterization by Kerr microscopy and Brillouin light scattering spectroscopy and x-ray studies were carried out to detect the dynamics in patterned magnetic antidot lattices. Here, we investigate Oersted-field driven spin dynamics in rectangular Ni$_\mathrm{80}$Fe$_\mathrm{20}$/Pt antidot lattices with different lattice parameters by electrical means {and compare them to micromagnetic simulations}. When the system is driven to resonance, a $dc$ voltage across the length of the sample is detected that changes its sign upon field reversal, which is in agreement with a rectification mechanism based on the inverse spin Hall effect. Furthermore, we show that the voltage output scales linearly with the applied microwave drive in the investigated range of powers. Our findings have direct implications on the development of engineered magnonics applications and devices.

\end{abstract}

\maketitle
Magnonic crystals, a new class of metamaterials with periodically modulated magnetic properties, have emerged as key building blocks in magnonics \cite{Kruglyak}. The realization of spin-wave filters \cite{Kim}, phase shifters \cite{Zhu}, interferometers \cite{Podbielski}, spin-wave logic devices \cite{Schneider} and grating couplers \cite{Sklenar_APL_2012} has been demonstrated and it is possible to tune the magnonic properties as desired by engineering the magnetic properties of the magnonic crystal. In this regard, ferromagnetic antidot lattices are prototypes of magnonic crystals. The periodicity, dimensions, shape and material of an antidot lattice dictate the spin-wave frequencies and their spatial distribution. These characteristics are influenced by inhomogeneities of internal magnetic fields in lattices with larger periods, whereas at smaller periods exchange fields play an important role \cite{Mandal}. Since those parameters can easily be tuned by designing the pattern and choosing the proper material, antidot lattices are of fundamental importance in magnonics. 

\begin{figure}[b]
\includegraphics[width=1\columnwidth]{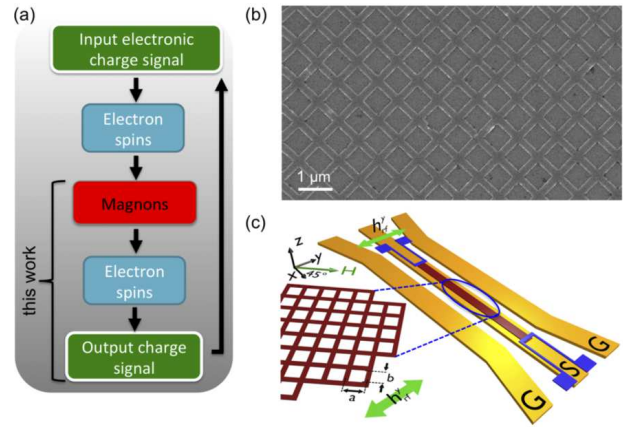}
\caption{\label{Fig1} (Color online) (a) Pathway for signal processing from an electronic to a magnonic signal and back. Signal processing can be achieved by magnons. (b) Example of a scanning electron microscopy image; here: antidot lattice $B$ with $a=845$~nm and $b=585$~nm. (c) Schematic of the experimental setup. The antidot lattice is oriented at 45$^\circ$ with respect to the signal line $S$. The dimensions $a$ and $b$ are given in Tab.~\ref{table}.} 
\end{figure}

Spin dynamics in antidot lattices were investigated by numerous resonance measurements  \cite{Neusser_PRL_2010,Pechan_JAP_2005,Sklenar_APL_2013}, by x-ray spectroscopy \cite{Graefe_Nanotech_2015} and by optical techniques such as Kerr microscopy and Brillouin light scattering spectroscopy \cite{Neusser_PRL_2010,Pechan_JAP_2005,Gubbiotti_APL_2015,Novosad_PRB_2002}. In order to utilize antidot lattices in real magnonic applications, however, it is desirable to integrate them in conventional $dc$ electronic devices and circuitries. The optimal signal processing pathway is \cite{Cornelissen}: input electronic charge signal $\rightarrow$ spin current signal $\rightarrow$ magnonic signal $\rightarrow$ spin current signal $\rightarrow$ output electronic charge signal, see Fig.~\ref{Fig1}(a). Signal processing and transfer can be realized by magnons and ultimately it would be possible to harness the unique and controllable magnon characteristics of the antidot lattice for real applications.

On one hand, spin Hall effects \cite{Dyakonov_JETP_1971,Dyakonov_PLA_1971} have been proven to be excellent candidates for the interconversion between electronic charge and spin currents \cite{Hoffmann}, and on the other hand, spin pumping \cite{Tserkovnyak_PRL_2002,Czeschka_PRL_2011,Mosendz_PRB_2010,Jungfleisch_PRB_2015,Jungfleisch_APL_2011} and spin-transfer torque \cite{Slonczewski_JMMM_1996,Berger_PRB_1996} effects are important methods to transform magnonic signals to spin currents and vice vesa \cite{Kajiwara,Zhang_JAP_2015,Sklenar_PRB_2015,Jungfleisch_SPIN_2015}. We will focus here on the detection side of the conversion pathway: the transformation of a magnonic signal (spin dynamics) into an electronic $dc$ charge signal, see Fig.~\ref{Fig1}(a).

In this Letter, we investigate spin dynamics in rectangular antidot lattices made of bilayers consisting of a ferromagnetic Ni$_\mathrm{80}$Fe$_\mathrm{20}$ layer (permalloy, Py) and a Pt capping layer with varying dimensions and periodicity. In contrast to conventional resonance experiments, where dynamics is driven by a microwave signal and the response of the magnetic systems is quantified by an absorption method, we use a rectification mechanism based on pure $dc$ techniques. Spin dynamics in the Py layer are induced by a microwave driven Oersted field and detected by {$dc$ voltage that is explained by a rectification based on the inverse spin Hall effect (ISHE) in the Pt.} It is shown that various modes determined by the design of the pattern contribute to the $dc$ voltage proving the possibility to utilize engineered antidot lattices for information processing/transport integrated in $dc$ circuitries. {The mode spectrum is confirmed by micromagnetic simulations.} Power dependent measurements confirm a linear response of the antidot lattice, which is desirable for the development of electronic devices.

The samples were fabricated in the following fashion: In a first step, $dc$ leads were fabricated by magnetron sputtering and photolithography on intrinsic Si substrates with a 300 nm thick thermally grown SiO$_2$ layer. The antidot lattices of various dimensions were then written by electron beam lithography (see Tab.~\ref{table}). A double layer positive resist of PMMA was spin-coated prior to electron beam exposure. After exposure and development, 15 nm-thick permalloy and 5 nm-thick Pt layers were deposited using electron beam evaporation at rates of \textless 0.3 \AA/s without breaking the vacuum. The resist was lifted-off in acetone. Figure~\ref{Fig1}(b) shows a typical scanning electron microscopy image (SEM), for lattice $B$, see Tab.~\ref{table}. The antidot lattices cover an area of approximately $800\times 20~ \mu$m$^2$  in total and the lateral dimensions of each investigated antidot lattice is summarized in Tab.~\ref{table}.  In a subsequent step a $50~\Omega$-matched coplanar waveguide (CPW) made of Ti/Au (3 nm/150 nm) was fabricated by magnetron sputtering and photolithography. The antidot lattice and the CPW were separated by a $80$~nm-thick MgO layer to avoid any electrical contact and the $dc$ leads are kept between the central line and the ground plate within the CPW in order to minimize inductively coupled currents in the sample.

\begin{table}[t]
\label{table} % is used to refer this table in the text
\caption{ \label{table} Overview of the investigated antidot lattices. Py thickness: 15~nm, Pt thickness: 5~nm.} % title of Table
\centering  % used for centering table

\begin{tabular}{c  c  c} % centered columns (4 columns)

\hline     \hline               %inserts double horizontal lines
Antidot lattice &  lattice constant $a$ (nm) & hole width $b$ (nm)\\ [0.5ex] % inserts table 
%heading
\hline                     % inserts single horizontal line
$A$ & 755 & 519  \\ % inserting body of the table
$B$ & 845 & 585 \\
$C$ & 942 & 713 \\     
\hline \hline%inserts single line
\end{tabular}

\end{table}

\begin{figure}[h]
\includegraphics[width=1\columnwidth]{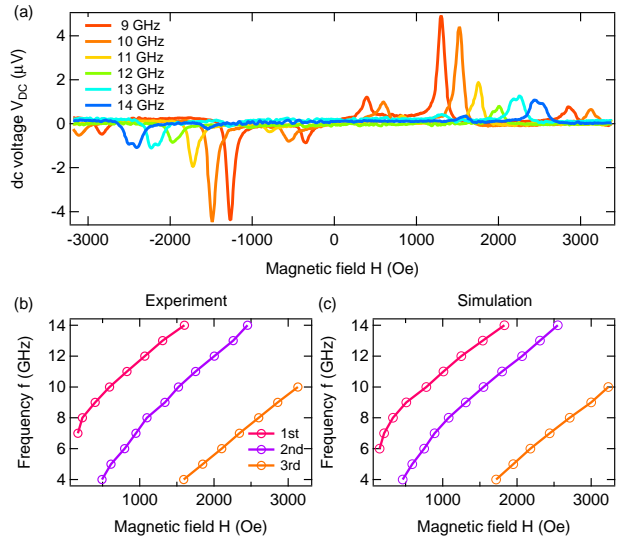}
\caption{\label{Fig2} (Color online) (a) Typical $dc$ voltage spectrum (here: antidot lattice $B$) measured at an applied power of $+15$~dBm. The low frequency spectra are omitted to provide better readability. The resonance signals show a mostly symmetric Lorentzian lineshape and change their polarity upon field reversal. (b) Frequency vs. field relation extracted from the spectrum shown in Fig.~\ref{Fig2}(a) for the three modes that can be identified clearly. {(c) Simulated frequency vs. field relation, here: antidot lattice $B$.}} 
\end{figure}

Figure~\ref{Fig1}(c) illustrates the experimental setup and the measurement configuration. The microwave driven Oersted field (in $y$-direction) is aligned at 45$^\circ$ to the external magnetic field [in $(1,1)$-direction; CPW and external field are oriented at an angle of 45$^\circ$], whereas the square antidot lattice is oriented parallel to it. Magnetization dynamics is excited by the Oersted field generated by a microwave current in the CPW. We use an amplitude modulation of the signal generator and lock-in technique to detect the $dc$ voltage output. For a particular measurement, the $rf$ power and frequency are kept constant (power range +10 to +18~dBm, frequency range: 4 to 14~GHz) while the external magnetic field is swept and the $dc$ output is recorded. When the system is driven to resonance, the magnetization precession in the Py layer generates a spin accumulation at the Py/Pt interface that diffuses into the Pt layer, a phenomenon that is known as spin pumping effect \cite{Tserkovnyak_PRL_2002}. This spin current gives rise to an electronic charge imbalance in the Pt layer due to the inverse spin Hall effect. The conversion from a spin- into charge current is described by:
\begin{equation}
\label{ISHE}
{\vec{J}_\mathrm{C}\propto\theta_\mathrm{SH}\vec{J}_\mathrm{S}\times\vec{\sigma},}
\end{equation}
where $\vec{J}_\mathrm{C}$ is the charge-current density, ${\theta}_\mathrm{SH}$ the spin Hall angle that describes the efficiency of the conversion, $\vec{J}_\mathrm{S}$ is the spin-current density and $\vec{\sigma}$ is the spin polarization vector.

\begin{figure}[t]
\includegraphics[width=1\columnwidth]{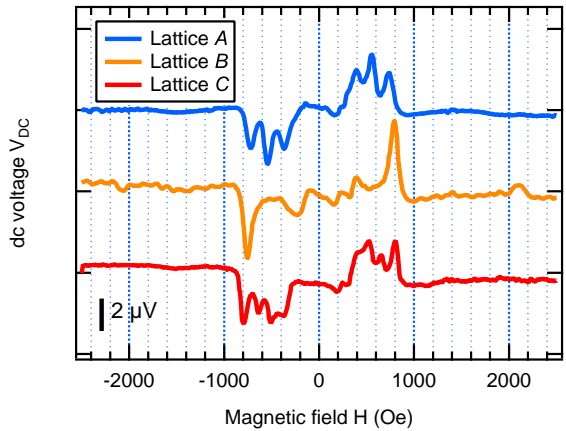}
\caption{\label{Fig3} (Color online) Comparison of the voltage spectra of the different antidot lattices at a fixed excitation frequency of 6 GHz and microwave power of +15~dBm. {The lattices feature the following stadium widths $(a-b)$ and whole width $b$. A: $(a-b)=236$~nm, $b=519$~nm, B: $(a-b)=260$~nm, $b=585$~nm, C: $(a-b)=229$~nm, $b=713$~nm}.} 
\end{figure}

The investigation of different mode spectra in antidot lattices and the examination of the underlying effects is very intriguing and subject of many studies in magnonics. However, we focus here on the fact that these well known resonances can indeed be detected by pure $dc$ electrical means.
A typical bi-polar spectrum recorded at an applied power of $+15$~dBm is shown in Fig.~\ref{Fig2}(a) (here shown: antidot lattice $B$). We clearly observe distinct modes in the $dc$ spectra at particular frequency--field values. The low-frequency modes are not shown in Fig.~\ref{Fig2}(a) to increase the readability of the viewgraph. 
{As is apparent from the figure, the modes exhibit a mostly symmetric Lorentzian lineshape. The most likely source for this behavior is a spin-to-charge current conversion due to the ISHE. For an unstructured sample oriented in the same way as here or if there was a substantial phase shift between the $rf$ current and the magnetization oscillation present, we would expect to observe a larger antisymmetric contribution if the signal was dominantly generated by a rectification due to the anisotropic magnetoresistance (AMR) \cite{Bai_PRL_2013}. Besides that, the polarity of the voltage signal changes sign when the magnetization direction is changed [Fig.~\ref{Fig2}(a)], which also suggests that the observed voltage is most likely due to the ISHE \cite{Bai_PRL_2013}, Eq.~(\ref{ISHE}). However, other effects such as spin rectification \cite{Bai_PRL_2013}, magnonic charge pumping \cite{Azevedo_PRB_2015,Ciccarelli_Nat_2015} or AMR  \cite{Bai_PRL_2013} cannot be completely ruled out. Independent of the rectification mechanism the experimental data unambiguously demonstrates an easy way to detect spin dynamics in magnonic crystals by pure $dc$ electrical means.}
%As is apparent from the figure, the modes exhibit a mostly symmetric Lorentzian lineshape that stems from a spin-to-charge current conversion due to the ISHE. In an unstructured sample oriented in the same way as here, we would expect to observe a larger antisymmetric contribution if the signal was dominantly generated by a rectification due to the anisotropic magnetoresistance (AMR) \cite{Bai_PRL_2013}. However, the lattice can be regarded as a series of resistors and since each resistor is aligned either along or perpendicular to the direction of the bias magnetic field, the AMR signal is minimized, whereas the ISHE contribution is maximum. Other effects such as spin rectification \cite{Bai_PRL_2013} or magnonic charge pumping \cite{Azevedo_PRB_2015,Ciccarelli_Nat_2015} might also play a role in structured samples. However, even more strikingly, the polarity of the voltage signal changes sign when the magnetization direction is changed [Fig.~\ref{Fig2}(a)], which is in agreement with a rectification based on ISHE \cite{Bai_PRL_2013}, Eq.~(\ref{ISHE}).
Furthermore, it is interesting to note that the magnitude of the detected $dc$ signal is comparable to that of unpatterned Py/Pt stripes \cite{Zhang_APL_2013}. In the spectrum shown in Fig.~\ref{Fig2}(a) we can clearly distinguish three different modes. We analyze their frequency--magnetic field dependence as shown in Fig.~\ref{Fig2}(b). As expected from %$f=\vert \gamma \vert/2\pi \sqrt{H(H+4\pi M_\mathrm{eff})}$, 
the Kittel equation, the resonance frequency increases with the externally applied magnetic field. {In order to corroborate the experimentally observed spectra, micromagnetic simulations were performed using mumax3 \cite{mumax3,simulations}. An exceptional good agreement between the theoretically expected and experimentally observed spectra is found. As in experiment three modes are found in the investigated magnetic field range, see Fig.~\ref{Fig2}(c).}

\begin{figure}[b]
\includegraphics[width=1\columnwidth]{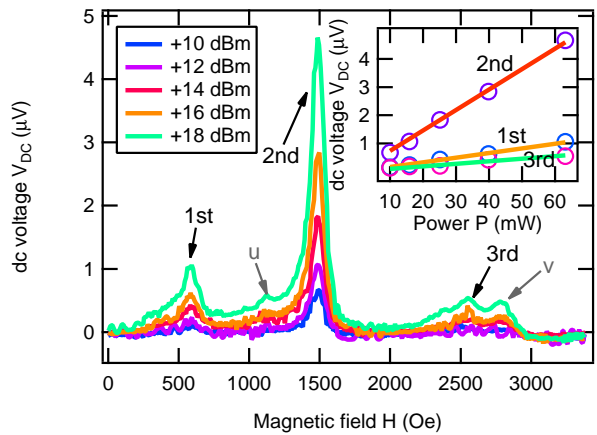}
\caption{\label{Fig4} (Color online) Voltage spectrum as function of the applied microwave power at an fixed excitation frequency of $10$~GHz (antidot lattice $A$). Three modes denoted as $1st$, $2nd$ and $3rd$, are detectable at all microwave powers. At higher powers additional modes (labeled as $u$ and $v$) emerge. The inset illustrates the linearity of the generated $dc$ output voltage with power; please note the linear scale of the power.} 
\end{figure}

Figure~\ref{Fig3} compares the $dc$ voltage spectrum of the investigated antidot lattices, excited at 6~GHz and +15~dBm. The different lattices show distinct features in the mode structure. The modes of lattice $A$ and $C$ lie closer together than those of $B$, leading to the conclusion that dipolar interactions emerging for narrower stadium widths ($a-b$) are the dominant tuning parameter here (see Tab.~\ref{table}). On the other hand, for the larger width $b$, a larger resonance field is observed. This can be understood as a decrease of demagnetization with increasing $b$ and, thus, an increase of the resonance field  at a particular excitation frequency \cite{Sklenar_APL_2013}. The magnitude of the detected voltage is basically independent of the lattice parameters, see Fig.~\ref{Fig3}. This demonstrates the possibility to tune the magnonic frequency characteristics as desired.

Next, we will focus on power-dependent studies. Linear response is a necessary requirement for the utilization of potential magnonic devices in electrical circuits. Therefore, we carried out microwave power dependent measurements. %Figure~\ref{Fig4} illustrates the power dependence of the spectrum of lattice $A$. 
As is apparent from the power dependence shown in Fig.~\ref{Fig4} that the $dc$ signals of all three modes increase with power. The magnitude of the output signal of each mode as a function of power is shown in the inset in Fig.~\ref{Fig4} (please note the linear scale; $P$ in mW). We can draw two important conclusions from this measurement. Firstly, we observe a linear response of the system: the output signal increases linearly with the drive in the investigated range of applied powers. This shows that we operate in the linear regime and no nonlinear dynamics is excited. Secondly, the first and third mode show approximately the same power dependence, whereas the second mode increases much faster with power. This might particularly be of interest for the development of magnonic devices where frequency/field dependent threshold output signals can be addressed by choosing an appropriate excitation power. Moreover, additional modes, which are not detectable at low powers emerge at higher excitation powers, e.g., at 1100 Oe and 2800 Oe for $f = 10$~GHz (labeled as $u$ and $v$ in Fig.~\ref{Fig4}). Since we observe a linear behavior of the three initial modes without any saturation, the modes $u$ and $v$ are not associated with the onset of nonlinear effects. The reason why they can only be detected at higher microwave powers is a better coupling of the driving fields to the magnetization.

In summary, we investigated the detection of spin dynamics in different square antidot lattices made of ferromagnetic metal - normal metal (Py/Pt) bilayers by means of spin pumping and inverse spin Hall effect. These investigations reveal that different modes characteristic for antidot lattices can be observed by $dc$ voltage output signals. Furthermore, we showed that the voltage signals of all modes scales linearly with the applied microwave power, yet the strongest mode shows the largest signal increase with power, which might directly affect the development of magnonic devices. Our studies demonstrate an easy way to investigate the properties of antidot lattices by a simple detection scheme and, even more importantly, the feasibility of an integration of antidot lattices as processing and transport devices in conventional electronics. However, in order to realize a full circle of conversion from an electronic to a magnonic signal and back to an electronic signal [Fig.~\ref{Fig1}(a)], the first conversion process by spin Hall and spin-transfer torque effect remains to be explored. Furthermore, our work suggests that a local detection and excitation of spin dynamics might be possible by covering only selective areas of the antidot lattice with Pt.

%\begin{acknowledgments}
This work was supported by the U.S. Department of Energy, Office of Science, Materials Science and Engineering Division. Lithography was carried out at the Center for Nanoscale Materials, an Office of Science user facility, which is supported by DOE, Office of Science, Basic Energy Science under Contract No. DE-AC02-06CH11357.
%\end{acknowledgments}


\begin{thebibliography}{19}
\bibitem{simulations}{We simulated the various antidot lattices $A$, $B$ and $C$ having the sane dimensions as in experiment (Tab.~\ref{table}) with $3\times3$ unit cells with periodical boundary conditions. Typical parameters for permalloy (saturation magnetization $M_\mathrm{S} = 700$~emu/cm$^3$, exchange constant $A = 13 \times 10^{-12}$~J/m and Gilbert damping constant $\alpha=0.01$) were used in the simulations. }
\end{thebibliography}
\end{document}